\documentclass[10pt,twocolumn,letterpaper]{article}

\usepackage{cvpr}
\usepackage{times}
\usepackage{epsfig}
\usepackage{graphicx}
\usepackage{amsmath}
\usepackage{amssymb}
\usepackage{epstopdf}
\usepackage{cite}
\usepackage{subfigure}
\usepackage{enumerate}
\usepackage{caption2}

\usepackage[pagebackref=true,breaklinks=true,letterpaper=true,colorlinks,bookmarks=false]{hyperref}

\cvprfinalcopy 

\def\cvprPaperID{****} 

\ifcvprfinal\fi
\begin{document}

\title{Gated Context Model with Embedded Priors for Deep Image Compression}

\author{
Haojie Liu, Tong Chen, Peiyao Guo, Qiu Shen, Zhan Ma \\
Vision Lab, Nanjing University
}

\maketitle

\begin{abstract}
 A deep image compression scheme is proposed in this paper, offering the state-of-the-art compression efficiency, against the traditional JPEG, JPEG2000, BPG and those popular learning based methodologies.  This is achieved by a novel conditional probably model with embedded priors which can accurately approximate the entropy rate for rate-distortion optimization. It utilizes three separable
   stacks to eliminate the blind spots in the receptive field for better probability prediction and computation reduction.  Those embedded priors can be further used to help the image reconstruction when fused with latent features, after passing through the proposed information compensation network (ICN). Residual learning with generalized divisive normalization (GDN) based activation is used in our encoder and decoder with fast convergence rate and efficient performance.  We have evaluated our model and other methods using rate-distortion criteria, where distortion is measured by multi-scale structural similarity (MS-SSIM). We have also discussed the impacts of various distortion metrics on the reconstructed image quality. Besides, a field study on perceptual quality is also given via a dedicated subjective assessment, to compare the efficiency of our proposed methods and other conventional image compression methods.
\end{abstract}

\begin{figure}[t]
   \centering
   \includegraphics[scale=0.18]{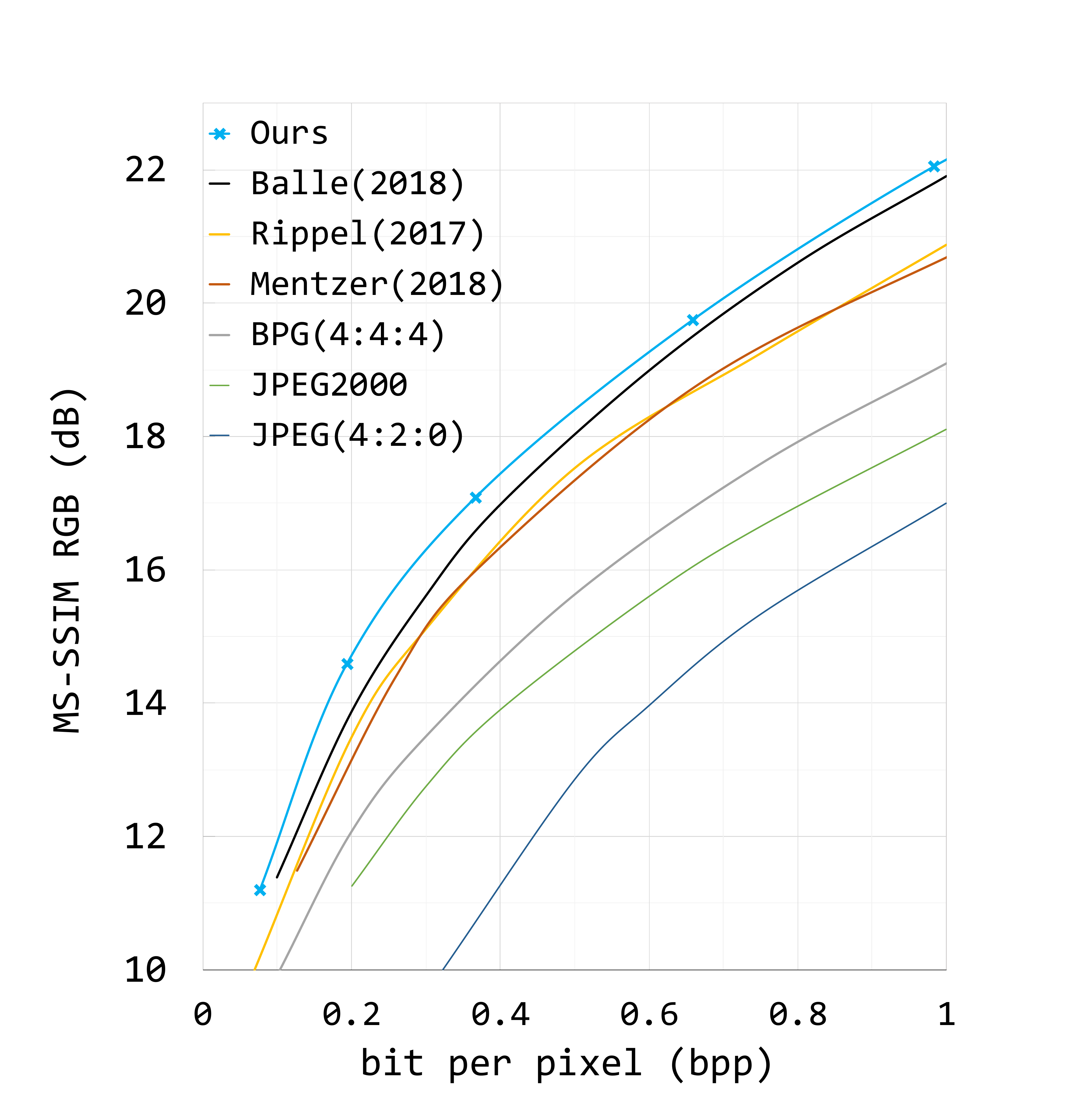}
   \caption{Our work achieves the state-of-art coding efficiency at all bitrates on \textbf{Kodak} dataset. Here we use $-10\log_{10}(1-d)$ to represent raw MS-SSIM ($d$) in dB scale.}
   \label{fig:rd_curve}
\end{figure}
\begin{figure*}[t]
     \centering
     \includegraphics[scale=0.34]{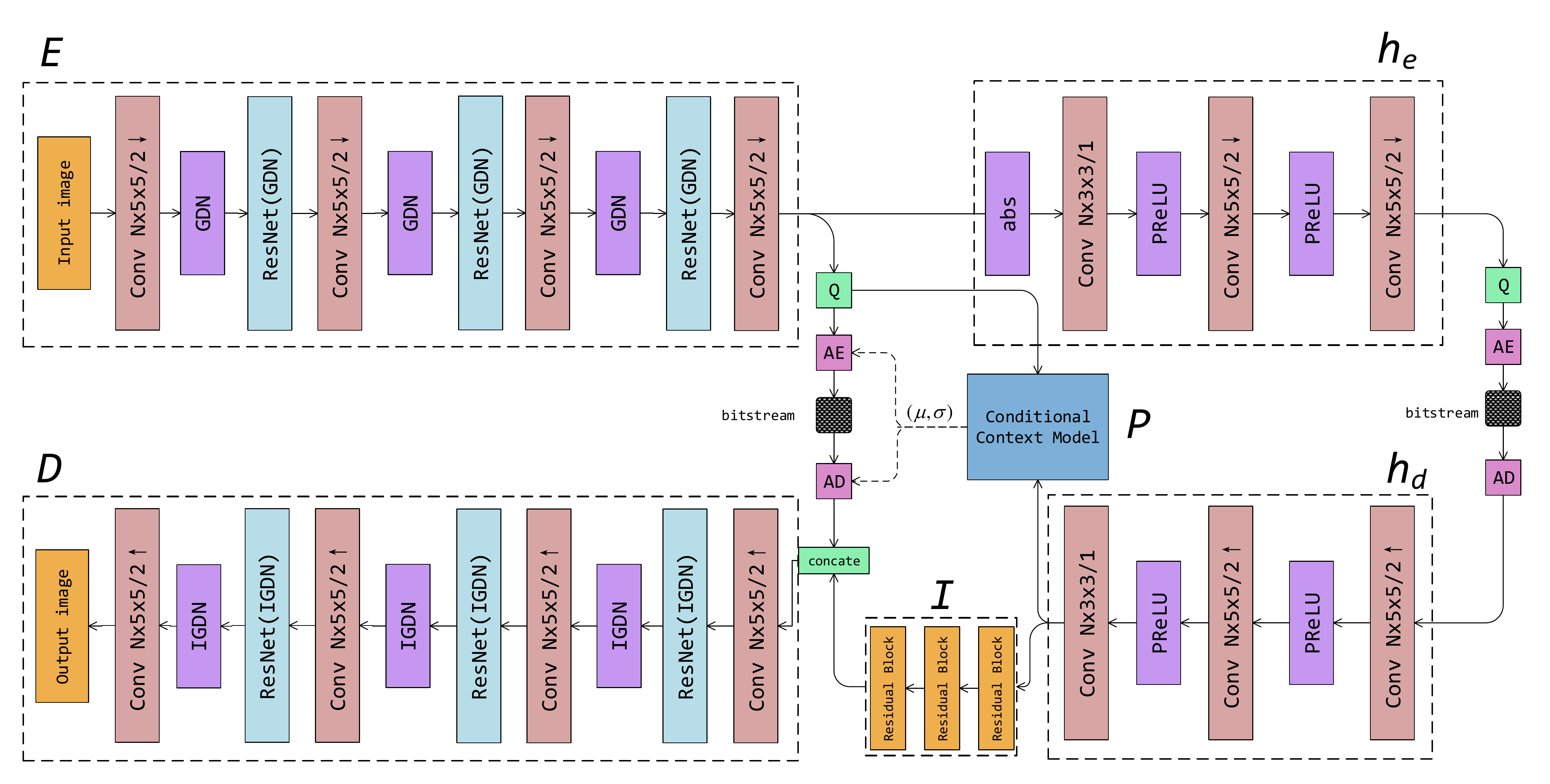}
     \caption{Illustration of our image compression framework. Each convolution layer is denoted as \#filters $\times$ width $\times$ height/scaling factor. We can set $N$ and $M$ to control the transform capacity and the upper bound of compression efficiency. All the residual nets (ResNets) use 3$\times$3 convolutions of stride 1 with add operations and remove the batch normalization (BN) layers. Activation function is replaced with generalized divisive normalization (GDN) in $E$ and $D$, while, in $h_e$ and $h_d$, we use parametric ReLU (PReLU) instead.}
     \label{fig:arch}
  \end{figure*}

\section{Introduction}
  The explosive growth of image/video data across the entire Internet poses a great challenge to network transmission and local storage, and puts forward higher demands for high-efficiency image compression. Conventionally, image compression methods, \eg, JPEG~\cite{wallace1992jpeg}, JPEG2000~\cite{taubman2012jpeg2000}, High-Efficiency Video Coding (HEVC) Intra Profile based BPG~\cite{bellard2016bpg}, \etc,  exploit and eliminate the redundancy via spatial prediction, transform and entropy coding.
   Nevertheless, these conventional methods can hardly break the performance bottleneck due to linear transforms with fixed basis, and limited number of prediction modes.
   On the other hand, machine learning based models~\cite{TodericiVJHMSC16,rippel2017real,mentzer2018conditional,balle2018variational} have led to a great success in image compression because of learned high-effective nonlinear transform and accurate entropy rate modeling.
   These learning based methods utilize the basic autoencoders to transform the image into
   compressible representations at the bottleneck layer followed by quantization and entropy coding to generate the binary stream.

   Usually, rate-distortion optimization (RDO) is used for image/video compression~\cite{Gary_RDO} to achieve the least distortion for a given rate constraint. Here, {\it how to accurately estimate entropy rate is the key}. For instance, 
   Mentzer \etal~\cite{mentzer2018conditional} proposed a 3D conditional probability model to approximate the entropy rate and applied an importance map for adapative spatial bits allocation. Inspired by the variational autoencoders (VAEs), Ball\'e \etal~\cite{balle2018variational} introduced hyperpriors as side information to construct a Gaussian distribution model for entropy probability estimation.
   Instead, in this work we introduce a {\it conditional context model} to produce appropriate probability density distribution (p.d.f) for better entropy rate estimation,
  where a gate mechanism is used to eliminate the blind spots in the receptive field caused by masked convolutions at deeper network layer, and a hyperprior is embedded as a latent vector $z$ to predict the conditional p.d.f $p(x|z)$ as a scale mixture of Gaussian distribution.
  We further transform the hyperprior $z$ using an information compensation network (ICN) to jointly reconstruct the image by concatenating with the compressed latent representation together.
  In addition, we use deep residual learning (ResNet)~\cite{he2016deep} together with the generalized divisive normalization (GDN) activation to construct latent features efficiently.

  Our method has demonstrated the state-of-art compression efficiency over the entire Kodak database~\cite{kodark}, when evaluated using the multi scale structural similarity (MS-SSIM) and actual bitrate as shown in Fig.~\ref{fig:rd_curve} in comparison with Ball\'e(2018) \etal~\cite{balle2018variational}, Rippel(2017) \etal~\cite{rippel2017real}, Mentzer(2018) \etal~\cite{mentzer2018conditional}, BPG~\cite{bellard2016bpg}, JPEG2000~\cite{taubman2012jpeg2000} and JPEG~\cite{wallace1992jpeg}. We also investigate the impacts of various distortion metrics (i.e., SSIM, MSE, VGG extracted feature-based distortion, \etc) on the rate-distortion efficiency. An additional subjective quality assessment is also provided to further study the efficiency of proposed method.


\section{Related Work} \label{sec:related_work}

Deep neural networks (DNN) based image compression generally depends on the well-known autoencoders and recurrent neural networks (RNN). These deep learning based methods are typically optimized using {\it back propagation} that requires all the steps differentiable. But the quatization operation usually does not meet this criteria. Thus, it is vital to implement an approximation process to replace conventional quantization (\eg, rounding). Theis \etal~\cite{theis2017lossy} proposed to replace the direct derivative using the derivative of the expectation, achieving an identity gradient passing. Ball\'e \etal~\cite{balle2016end} added uniform noise to simulate the quantization approximately in training stage and replaced it with direct rounding operation at inference step. Mentzer \etal~\cite{mentzer2018conditional} applied a soft-to-hard quantization  using nearest neighbor assignments and computed the gradients relying on soft quantization~\cite{agustsson2017soft}.

 Lossy image compression is mainly to perform the RDO, i.e.,  $J = R+\lambda D$, with $R$ for the rate and $D$ for the distortion such as mean square error (MSE) or MS-SSIM.
 $\lambda$ is used to balance the rate and distortion trade-off for compressive models at various compression ratios.
 At the very beginning, MSE was used for RDO~\cite{TodericiVJHMSC16,li2017learning,theis2017lossy,balle2016end}, which often incurred  unpleasant visual experience particularly at a very low bitrate.
 MS-SSIM was then introduced as the distortion measurement because of its better correlation with human vision system (HVS)~\cite{rippel2017real,mentzer2018conditional,balle2018variational}. In the meantime, generative adversarial networks (GAN) and feature domain loss based distortion measurements in RDO process ~\cite{ledig2017photo,rippel2017real,agustsson2018generative} can produce extremely impressive reconstructions,
 but comes with fake texture and structure compared with reference image.

 PixelRNNs and PixelCNNs~\cite{oord2016pixel} utilize the historical data (i.e., pixel intensity, pixel p.d.f) to predict the information at current location with masked convolutions. PixelRNNs generally give better performance but PixelCNNs present higher computational efficiency. Mentzer \etal~\cite{mentzer2018conditional} extended the 2D PixelCNNs to a 3D structure for compression entropy modeling. However PixelCNNs always suffer from the blind spots which will be enlarged as the growth of receptive field observed in~\cite{van2016conditional}.
Ball\'e~\cite{balle2018variational} then proposed to construct hyperpriors and calculate the variance for Gaussian distribution as p.d.f for entropy rate modeling.

 Besides, other methods such as pyramidal decomposition~\cite{rippel2017real} and generalized divisive normalization (GDN) based nonlinear transformation are introduced to improve the compression performance. GDN is proved to be more effective than other nonlinear activations in learning based compression~\cite{balle2018efficient}. But it usually leads to unstable training using Adam optimization with a larger learning rate.

\section{Proposed Method}
We define our image compression framework in Fig.~\ref{fig:arch} using an encoder $E$, a decoder $D$, an information compensation network (ICN) $I$, a hyper encoder $h_e$, a hyper decoder $h_d$ and a conditional context model $P$ using the output of $h_d$ as embedded priors.
Given an image $x$, the encoder $E$ transform $x$ into the latent representation $y=E(x)$.
Then we add uniform noise to approximate the quantization operation $Q$ to generate the quantized representation $\hat{y}=Q(y)$.
A hyper encoder $h_e$ with quantization is introduced to transform $y$ into more compact representation $\hat{z}=Q(h_e(y))$ as side information.
We use an symmetric hyper decoder $h_d$ to construct the symbols $z_p=h_d(\hat{z})$ for  conditional probability modeling and image reconstruction via ICN $I$. More specifically, ICN generated information is then concatenated with $\hat{y}$ for the final reconstruction of $\hat{x}=D([\hat{y},I(z_p]))$.
Here [.] represents the channel concatenation function. A gated conditional context model
$P$ is applied to provide the mean $\mu$ and scale variance $\sigma$ jointly based on $z_p$ and $\hat{y}$ using $(\mu,\sigma)=P(\hat{y},z_p)$.
Note that $z_p$ is treated as embedded priors for both $P$ and $D$ to improve the image reconstruction and the conditional probability modeling during the training.

\begin{figure}[t]
   \centering
   \includegraphics[scale=0.3]{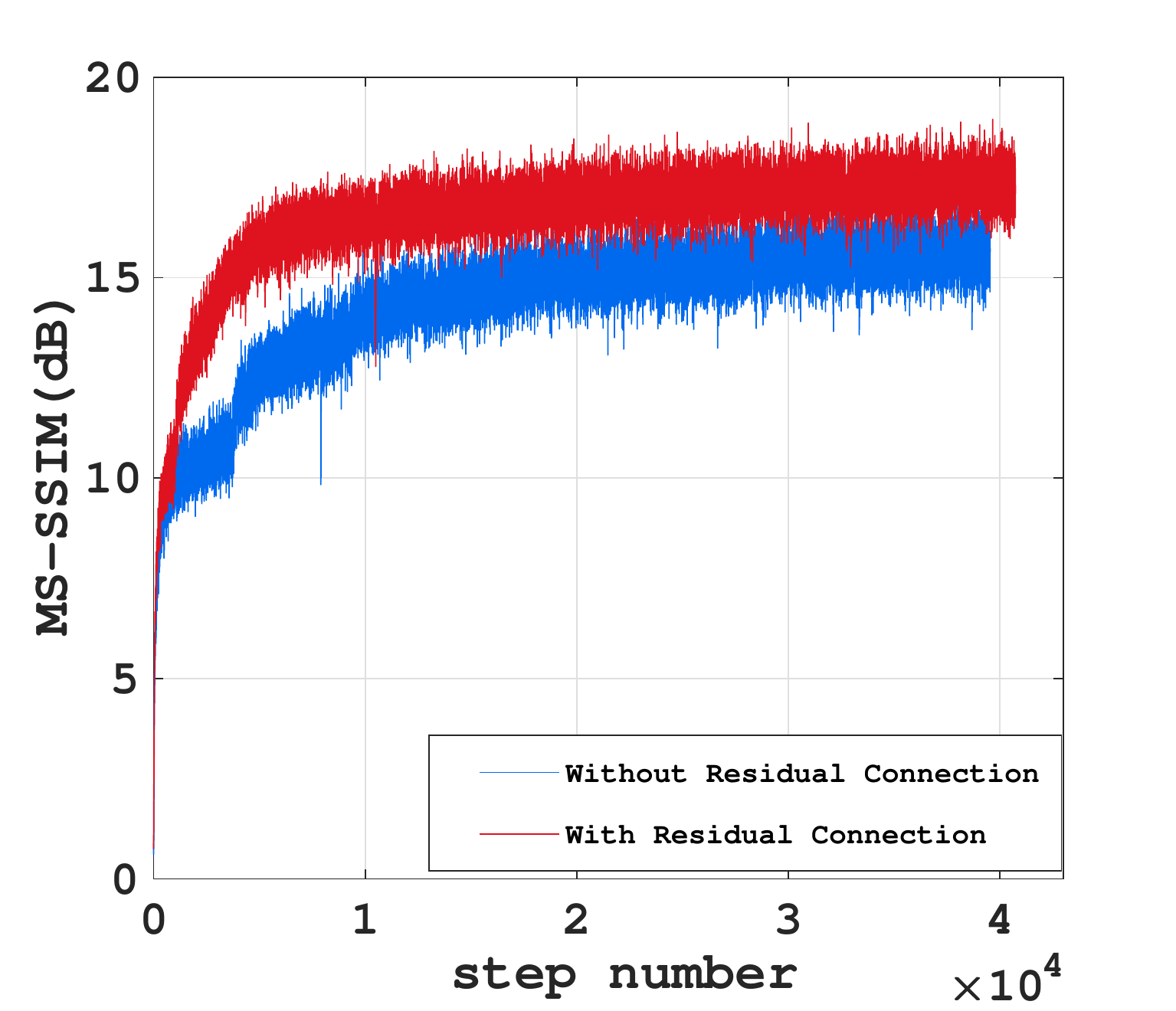}
   \caption{Convergence speed comparison: Residual learning with GDN can make the network converge much faster. For fair comparison, we conduct the tests for 10 times and keep other parameters as the same. We use adam optimizer with learning rate $\rho$ as $10^{-4}$. }
   \label{convergence_comparison}
\end{figure}

\subsection{Residual learning with GDN Activation}
GDN has been proved to be a good density model of images, and shows impressive performance compared with other nonlinear activation functions such as ReLU, leakyReLU and softplus used in compression networks~\cite{balle2018efficient}.
However with the growth of transform capacity and deeper architecture, the training becomes unstable using a larger learning rate. Even with a smaller learning rate, it often converges slowly and usually gets stuck to sub-optimal results. Klopp \etal~\cite{klopp2018learning} proposed a sparse GDN that uses ReLU in GDN function, to make sparse activation for better results than those reported in~\cite{balle2018variational}. Instead, inspired by high efficiency of ResNet~\cite{he2016deep}, we use residual learning framework but replace the conventional ReLU activation with the GDN, yielding very effective performance for training. It is shown in Fig.~\ref{convergence_comparison} that for the same number of layers used in our framework, residual connections can achieve 4$\times$ faster of convergence rate and marginal performance improvement.

\subsection{Entropy Rate Modeling}

We use different density models for $\hat{y}$ and $\hat{z}$.
We model the priors $\hat{z}$ using a non-parametric, fully factorized density model following~\cite{balle2018variational}. We convolve it with a standard uniform density to get $p_{\hat{z}|\psi}$,
\begin{equation}
  p_{\hat{z}|\psi}(\hat{z}|\psi) = {\prod_i} ( p_{z_i|\psi^{(i)}}(\psi^{(i)})*\mathcal{U}(-\frac{1}{2},\frac{1}{2})) (\hat{z}_i),
  \label{Eq3}
 \end{equation}
where $\psi^{(i)}$ represents the parameters of each univariate distribution $p_{\hat{z}|\psi^{(i)}}$.

For $\hat{y}$, each element $\hat{y}_i$ can be modeled as a Gaussian distribution,
\begin{equation}
  p_{\hat{y}|\hat{z}}(\hat{y}|\hat{z}) = {\prod_i} (\mathcal{N(\mu,\sigma)} *\mathcal{U}(-\frac{1}{2},\frac{1}{2})) (\hat{y}_i),
  \label{Eq3}
 \end{equation}
where its $\mu$ and $\sigma$ are predicted by $P(\hat{y},z_p)$. We can simply use the cumulative distribution function (CDF) to calculate the probability of each symbol for Gaussian distribution. Note that $P$ represents the gated conditional context model that takes the $z_p$ as a hidden vector to jointly derive $\mu$ and $\sigma$ with $\hat{y}$.

\subsubsection{Gated 3D Context Model}
\begin{figure}[b]
   \centering
   \includegraphics[scale=0.5]{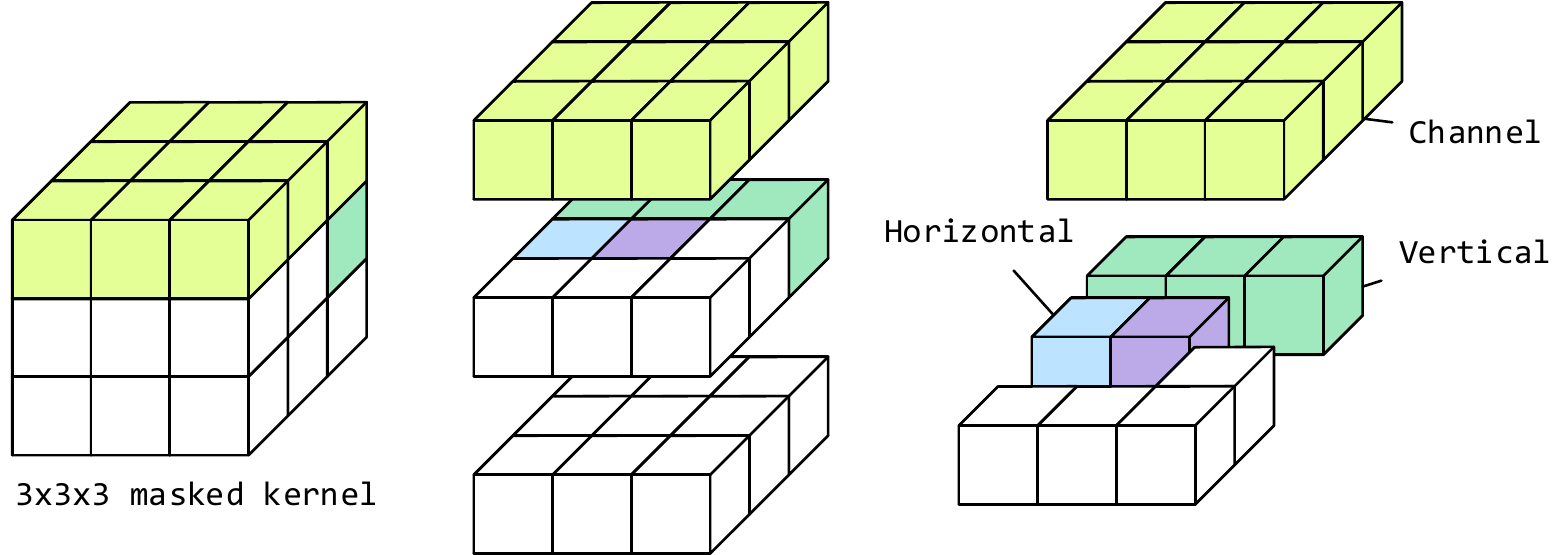}
   \caption{Illustration of masked 3D kernel. For a 3$\times$3$\times$3 convolution kernel, we just split the kernel into 3 separable stack called channel stack, vertical stack and horizontal stack. It can intuitively eliminate the blind spots caused by traditional masked PixelCNNs. }

   \label{fig:3 dimensional stacks}
\end{figure}

To predict $\mu$ and $\sigma$ of Gaussian distribution, we propose a PixelCNN~\cite{oord2016pixel} alike method for entropy rate modeling. Traditional PixelCNNs usually predict the pixels in a raster scan manner and estimate  next channel by combining former reconstructed and current channel information. Note that $x_i$ is the current pixel and the conditional distribution is represented as:
\begin{equation}
  p(x) = {\prod_i} p(x_i|x_1,...,x_{i-1}),
  \label{Eq4}
 \end{equation}
  Mentzer \etal~\cite{mentzer2018conditional} introduced a 3D probability model and proved to be more powerful than other traditional methods. However, if we make the context model larger for better prediction by adding more layers, it usually loses the information at the top-right corner. To solve such blind spot problems, we design a novel gated probability model inspired by~\cite{van2016conditional} to split the 3D convolution kernal   into 3 separable stacks, i.e., channel stack, vertical stack and horizontal stack as shown in Fig.~\ref{fig:3 dimensional stacks}.

\begin{figure}[t]
   \centering
   \includegraphics[scale=0.4]{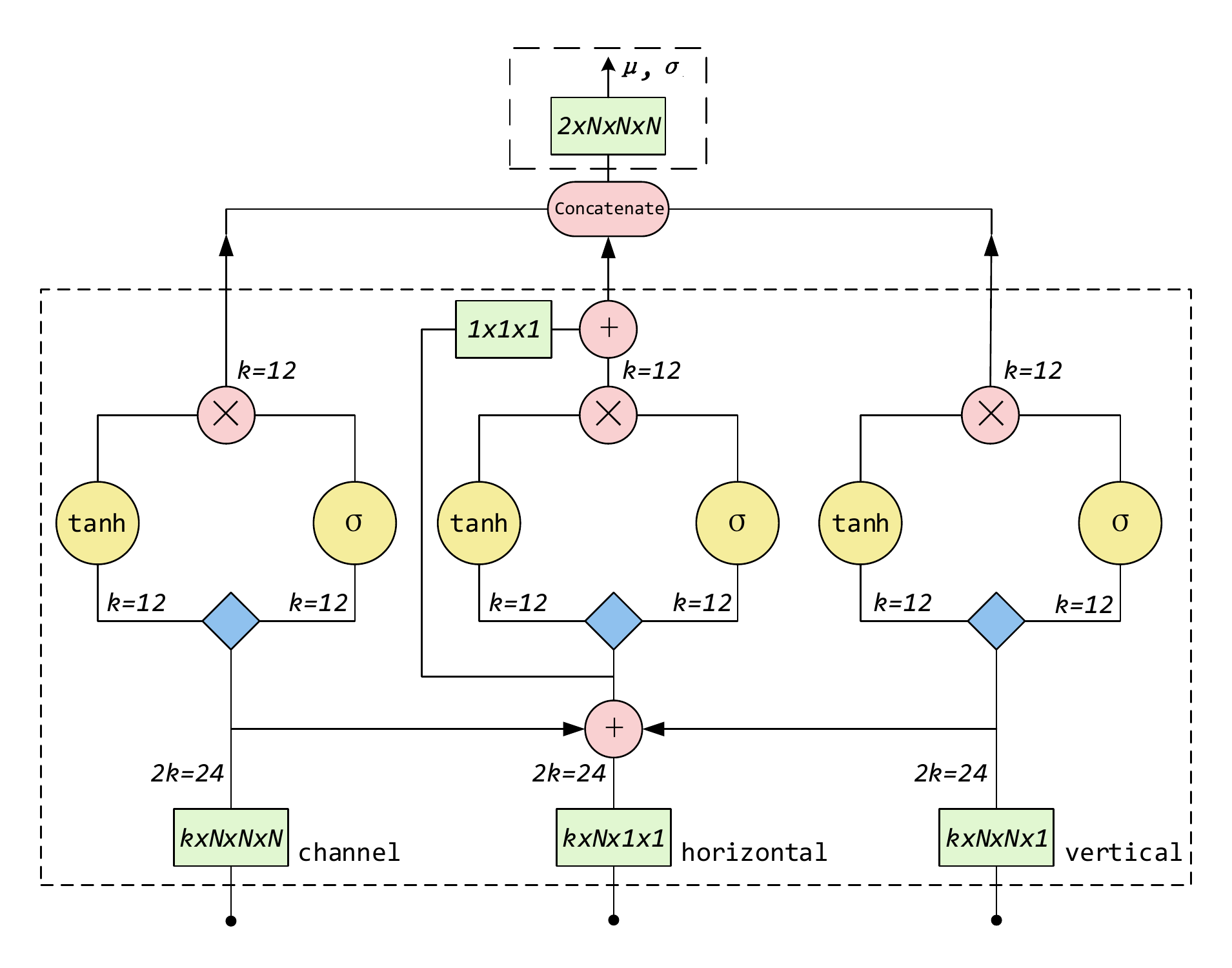}
   \caption{Illustration of the single layer of our gated context model. Here the green box denotes the 3D convolutions, the blue box denotes the split function and convolution, and the red box represents add and element-wise multiplication operations. $k$ represents the output dimension. We set $k$ to 12 in this paper. When it comes to the last layer, we concatenate 3D output and add a 1${\times}$1${\times}$1 convolution layer for information fusion to generate the final output. To make it computational efficient, we only apply a 3-layer context model. Note that we reshape the input to make $k$ be 1 for 3D convolutions and remove the residual connections in the first layer.}

   \label{the architecture of the gated model}
\end{figure}

   Therefore, we can access the entire neighbors of previous pixels in a 3D cubic. It is supposed that a ($n{\times}n{\times}n$) convolution kernel with a mask can be simply split into ($n{\times}n{\times}\frac{{\lceil}n{\rceil}}{2}$), ($n{\times}\frac{{\lceil}n{\rceil}}{2}{\times}1$) and ($\frac{{\lceil}n{\rceil}}{2}{\times}1{\times}1$)
convolutions via appropriately padding and cropping. It is also proven to be a very effective way to increase the parallelization and improve the prediction performance of context model. Each output of stack can be expressed as:
\begin{equation}
  v = \tanh(W_{k_1}*a){\odot}{\sigma}(W_{k_2}*a),
  \label{Eq5}
 \end{equation}
where $a$ denotes the input of separable stacks, $v$ denotes the output and $W_k$ is the weight of convolution. It performs the similar way as gate mechanism which is widely used in LSTM~\cite{gers2001long} to model the complex interactions. In fact, residual connections between each stack are also applied to formulate our final gated model as shown in Fig.~\ref{the architecture of the gated model}.

\subsubsection{Conditional Generation}
    Conditional generation models commonly utilize the embedding as extra information to generate the output which is related to it. Conditional generative adversarial networks (CGAN)~\cite{mirza2014conditional} introduced a method to generate the images belonging to the relevant category with a caption. For sequential data, Mathieu \etal~\cite{mathieu2015deep} utilized the previous reconstructed frame as priors to combine with the generated frames for training the discriminator. We predict the current pixel using the former pixels and the hyperpriors produced by $h_d$ as additional conditions and rewrite Eq.~\eqref{Eq4} and~\eqref{Eq5} as:
\begin{figure*}[t]
   \centering
   \includegraphics[scale=0.095]{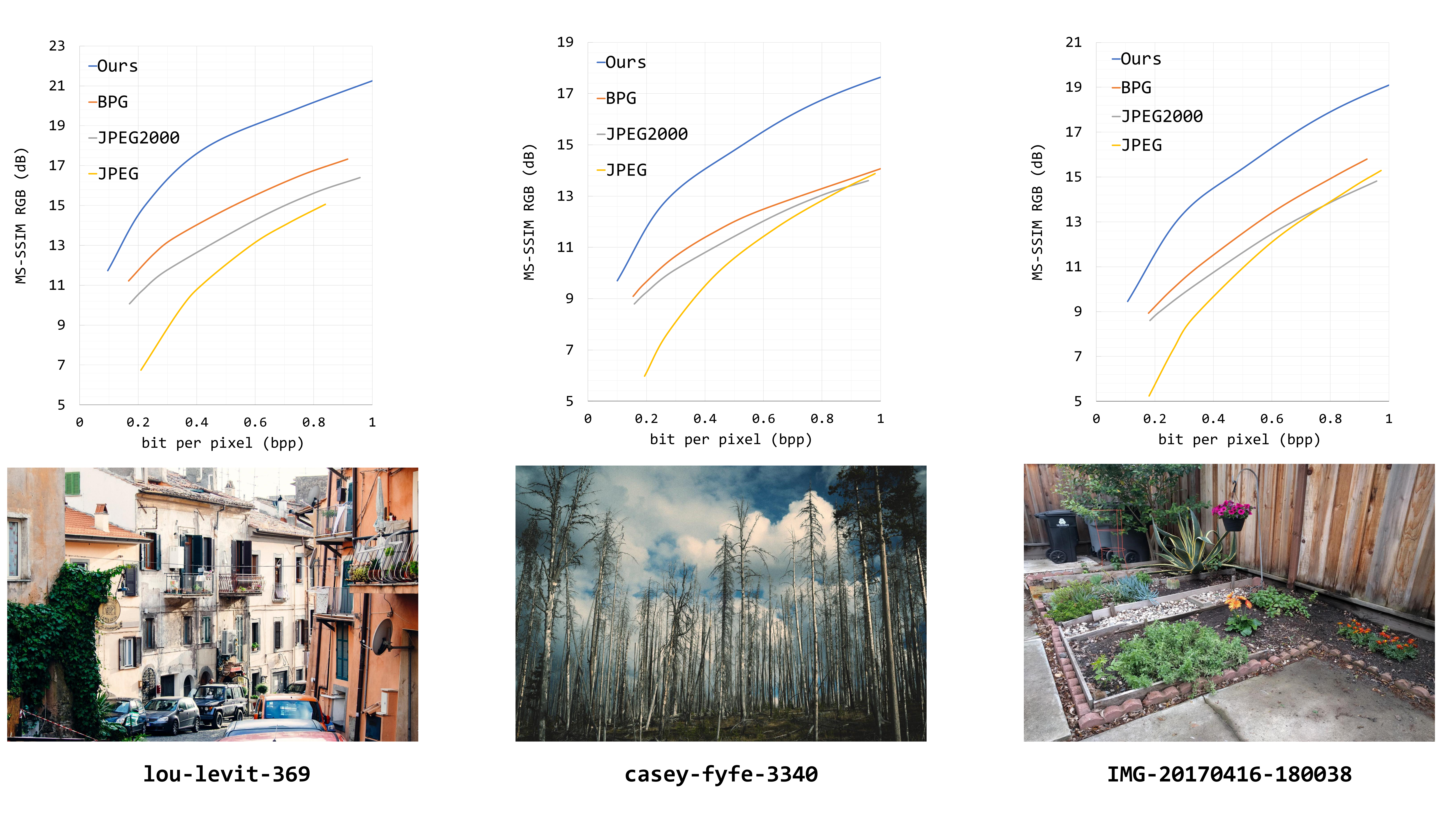}
   \caption{Compression performance of our method using three images from \textbf {CLIC}, compared with JPEG, JPEG2000 and BPG.}
   \label{fig:curve}
\end{figure*}

\begin{equation}
p(x) = {\prod_i} p(x_i|x_1,...,x_{i-1},{\textbf{h}}),
  \label{Eq6}
 \end{equation}
 \begin{equation}
   v = \tanh(W_{k_1}*a+V_{k_1}*{\textbf{h}}){\odot}{\sigma}(W_{k_2}*a+V_{k_2}*{\textbf{h}}),
   \label{Eq7}
  \end{equation}
where $\textbf{h}$ denotes the output of the hyper decoder $z_p$, $V_k$ is the 1$\times$1$\times$1 convolution kernel.

Eq.~\eqref{Eq7} describes a basic layer in our gated conditional context model.
We split the final output into $\mu$ and $\sigma$ as the same size of $\hat{y}$.
By calculating probability of each symbol $\hat{y}_i$, we can further get the entropy of $\hat{y}$:
\begin{equation}
  R_{\hat{y}} = - {\sum_i} {\log_2}(p_{\hat{y}_i|\hat{z}_i}(\hat{y}_i|\hat{z}_i)),
  \label{Eq8}
 \end{equation} while the entropy of $\hat{z}$ is calculated as the same:
 \begin{equation}
 R_{\hat{z}} = -{\sum_i}  {\log_2}(p_{\hat{z}_i|\psi^{(i)}}({\hat{z}_i}|\psi^{(i)}))
 \label{Eq9}
\end{equation}

\subsection{Information Compensation Network}

Ball\'e \etal~\cite{balle2018variational} proposed a hyper model as a probability model for entropy coding. During the training, the hyperpriors $\hat{z}$ consume a portion of bits that is referred as the side information cost.
We could consider the whole framework as a scalable compression system, where the essential problem is how to allocate the bits between latent features $\hat{y}$ and hyperpriors $\hat{z}$.
 While more bits are spent to $\hat{z}$,  an ICN is necessary to fully exploit the information and correlation contained in $\hat{z}$ for final image reconstruction.
 Although we use concatenation operation to fuse the information, an add operation can also be adapted at the bottom layer to treat $\hat{z}$ as residual information as presented in~\cite{TodericiVJHMSC16}. Overall, we provide a 3-layer modified residual network to achieve information reconstruction and pass-through. We remove Batch Normalization layers (BN) and replace the ReLU with PReLU in this study.

\section{Experimental Studies}
This section details the compression efficiency for our proposed framework. We set $N$ and $M$ to 192 and 128 to train the entire system in an end-to-end fashion.

\begin{figure*}[t]
   \centering
   \includegraphics[scale=0.32]{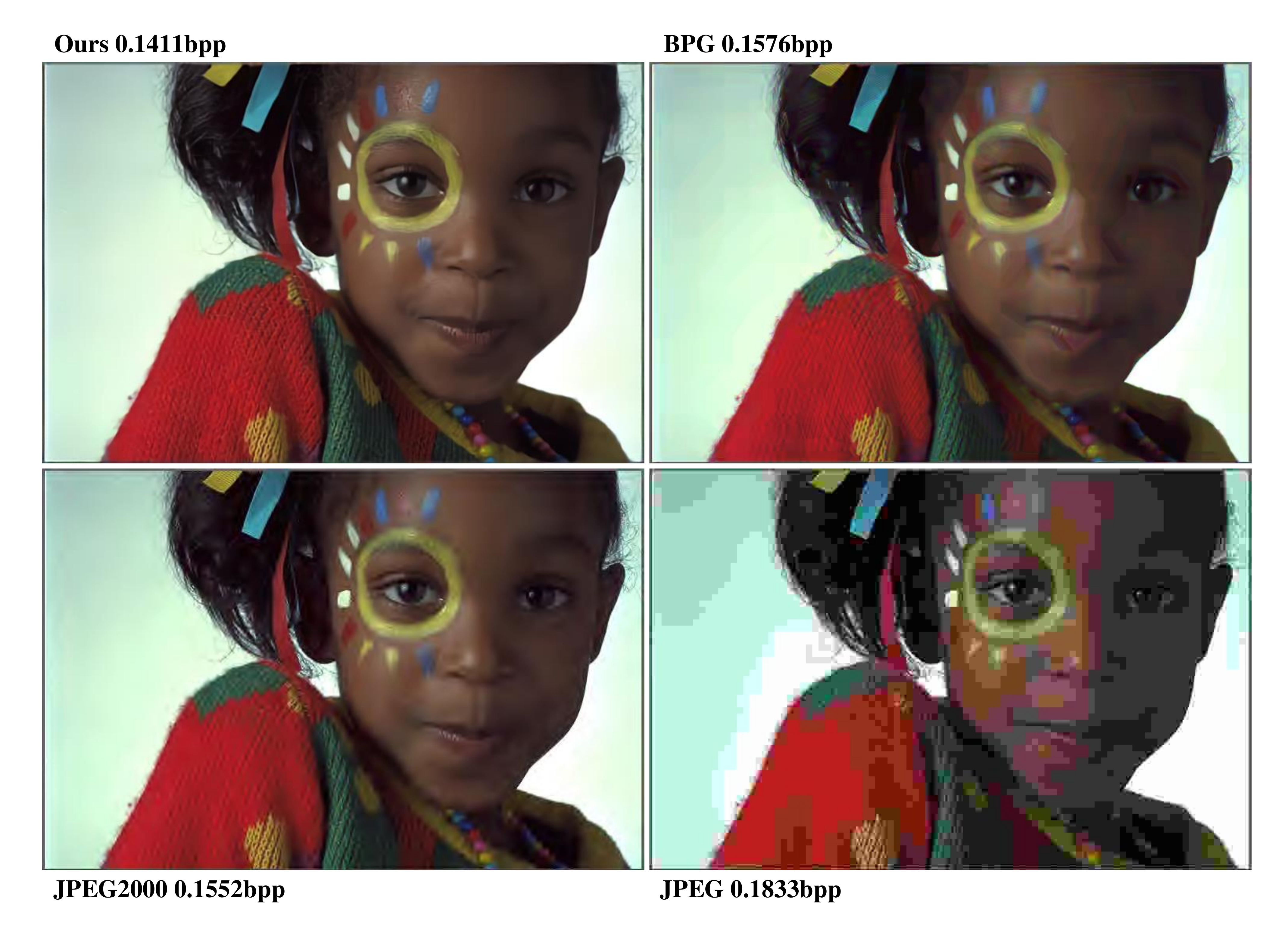}
   \caption{Snapshots of reconstructed ``kodim15'' that were compressed using various methods.}
   \label{fig:kodim15_comparison}
\end{figure*}

\textbf {Dataset:} We use \textbf{COCO}~\cite{lin2014microsoft} and \textbf{CLIC}~\cite{clic} training dataset to train our framework. We randomly resize the images and take 256 $\times$ 256 crops for preprocessing.  Furthermore, we test the results on the standard \textbf {Kodak} dataset~\cite{kodark} for fair comparison.

\textbf {RDO in Encoder Control} We choose MS-SSIM as our distortion metric $d$ and the loss function is described as:
\begin{equation}
L = {\lambda}(1-d)+R_y+R_z
\label{Eq10}
\end{equation}
where we set different $\lambda$ to achieve rate-distortion trade-off to generate several models for variable compression ratio.
In our experiment, we set $\lambda$ to 2, 8, 32, 128 and 384. We replaced the MSE with MS-SSIM~\cite{wang2003multiscale} because it exhibits better correlation with the subjective quality perceived by our HVS.
We then also study the impact of various distortion metrics on the compression efficiency, particularly for low bitrate range.

\textbf {Training:} All of the modules in our framework are trained together.
We set different learning rate $\rho$ for $E$, $D$, $h_{e}$, $h_{d}$ and $P$. For $E$, $D$, $h_{e}$, $h_{d}$, we use a $\rho$ of $10^{-4}$ and clip the value after 30 epoch to $10^{-5}$. For $P$, we use a $\rho$ of $5 {\times} 10^{-5}$.
Batch size is set to 64 and finally trained on 4-GPUs in parallel. To evaluate the results for different distortion metrics, we retrain our compression framework accordingly, and control the bitrate close to each other as much as possible.

\textbf{Performance:} The performance comparison measured on \textbf{Kodak} is shown in Fig.~\ref{fig:rd_curve}. Our method offers the state-of-the-art efficiency by outperforming the traditional codecs such as BPG, JPEG2000 and JPEG as well as those learning based methodologies reported in~\cite{mentzer2018conditional,rippel2017real,balle2018variational}. We also plot the R-D curves for three typical images in Fig.~\ref{fig:curve}, to further evident the superior performance of our proposed method over the traditional image codecs. In the mean time,  Fig.~\ref{fig:kodim15_comparison} visualizes the sample snapshots of reconstructed ``kodim15'' that were compressed at similar bit rate using our proposed method, BPG, JPEG and JPEG2000. We can clearly observe that our method have provided a better reconstruction with clean details and sharp edges. Besides, in the subsequent Section~\ref{sec:sqa}, we have given in-depth studies on the perceptual quality of compressed images via a dedicated subjective quality assessments.

\begin{figure*}[t]
   \centering
   \includegraphics[scale=0.2]{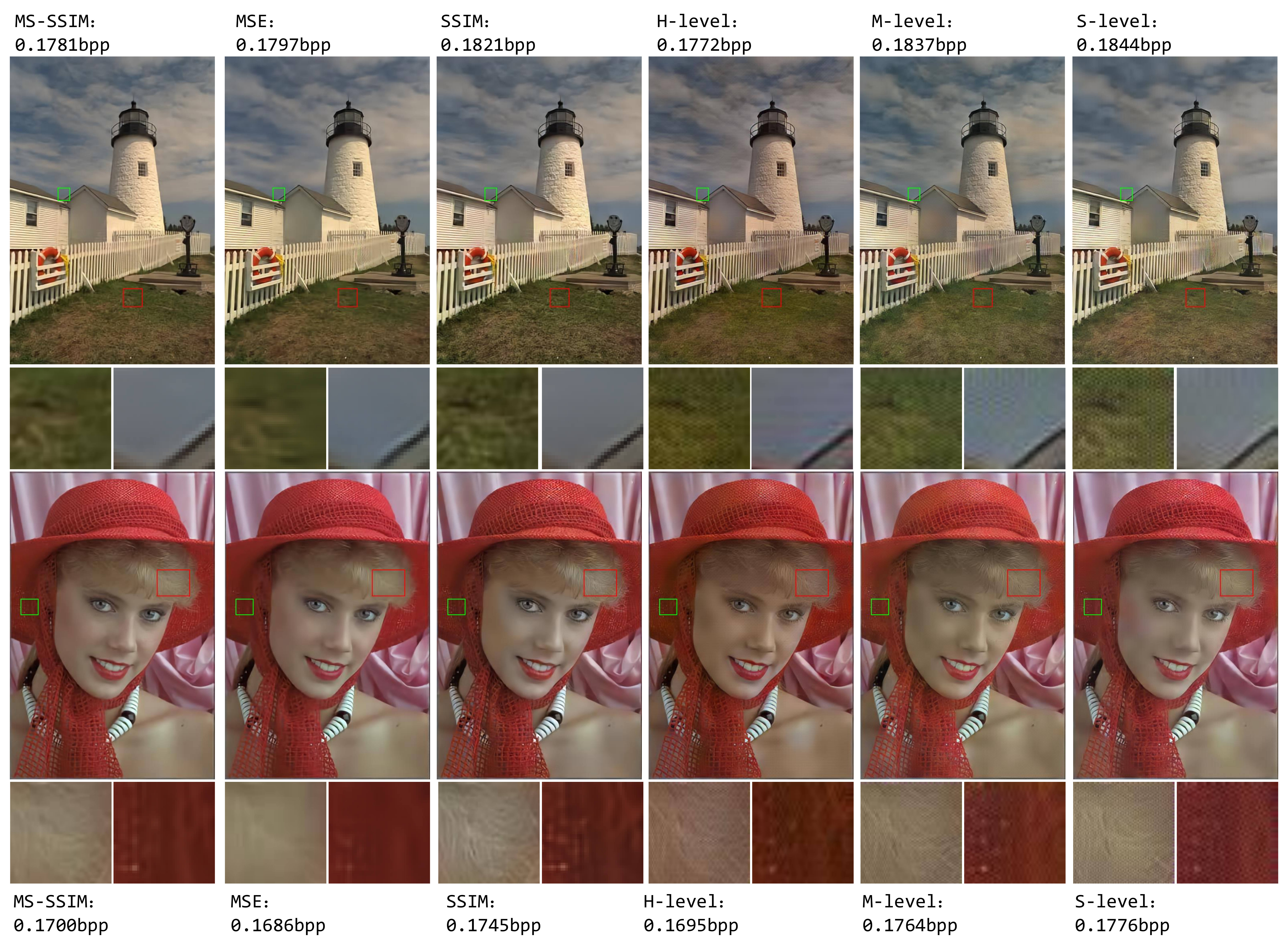}
   \caption{Comparison of different metrics: We control the bitrate under 0.2 bpp and optimize our compression framework using different loss functions. Here H-, M- and S-level denotes high-, mid- and shallow-level features based MSE.}
   \label{fig:kodim19_metric}
\end{figure*}

\begin{figure*}[h]
\centering
\subfigure[{Door $\protect\\$    SI =43.890} ]{\includegraphics[scale=0.076]{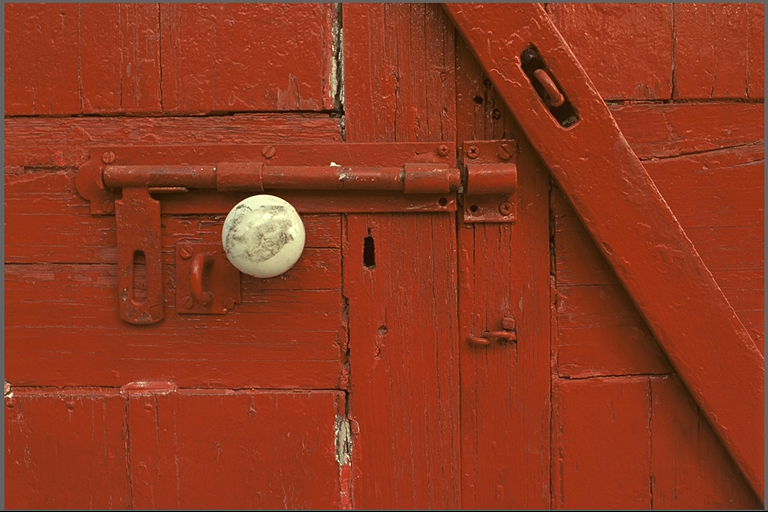}}
\subfigure[Motobike $\protect\\$    SI =108.495]{\includegraphics[scale=0.076]{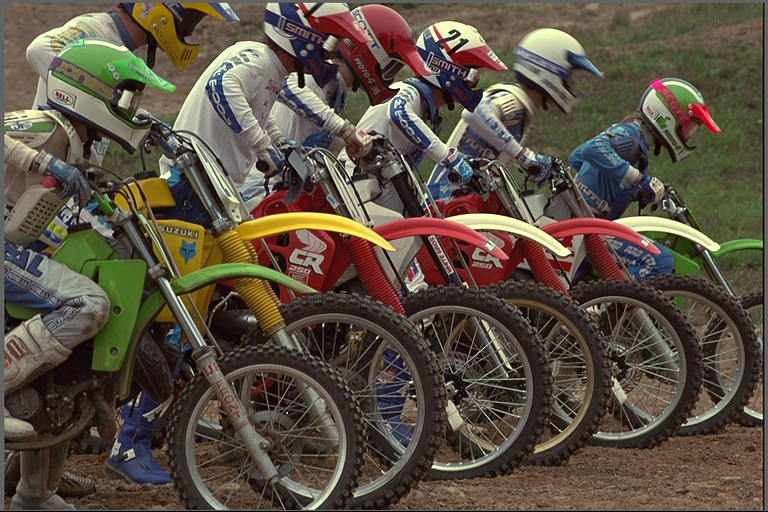}}
\subfigure[Apartment$\protect\\$    SI =136.290]{\includegraphics[scale=0.076]{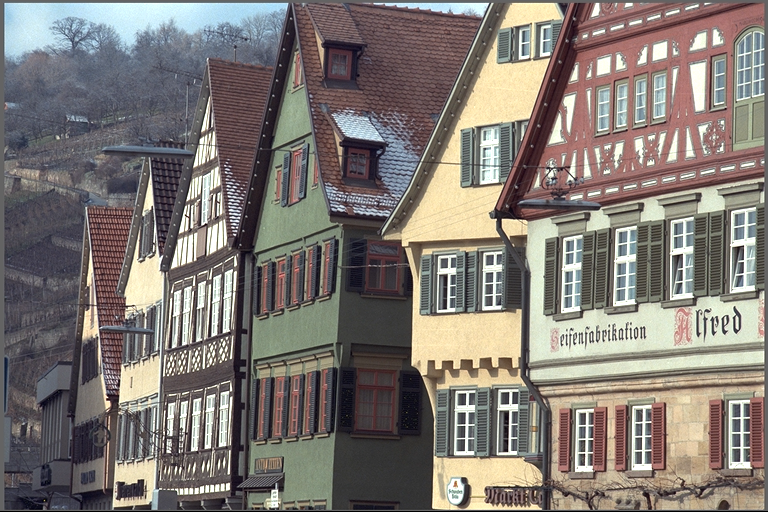}}
\subfigure[Girl$\protect\\$    SI = 57.005]{\includegraphics[scale=0.076]{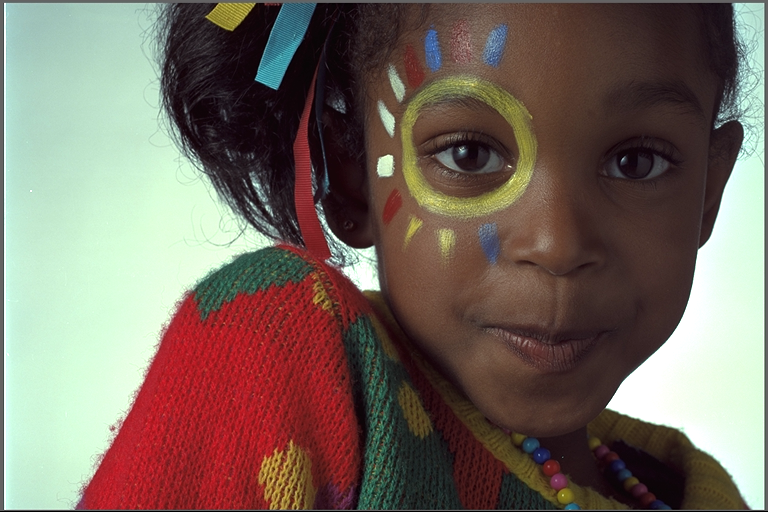}}
\subfigure[Boat$\protect\\$    SI =74.675]{\includegraphics[scale=0.076]{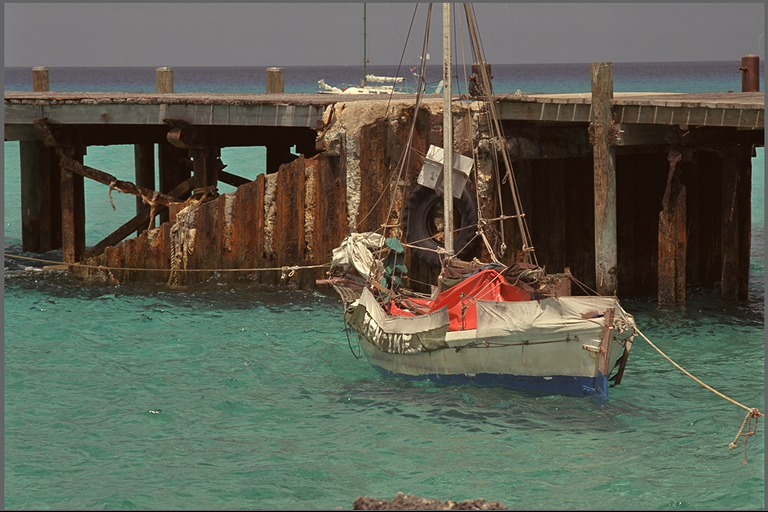}}
\subfigure[Lighthouse$\protect\\$    SI =84.368]{\includegraphics[scale=0.076]{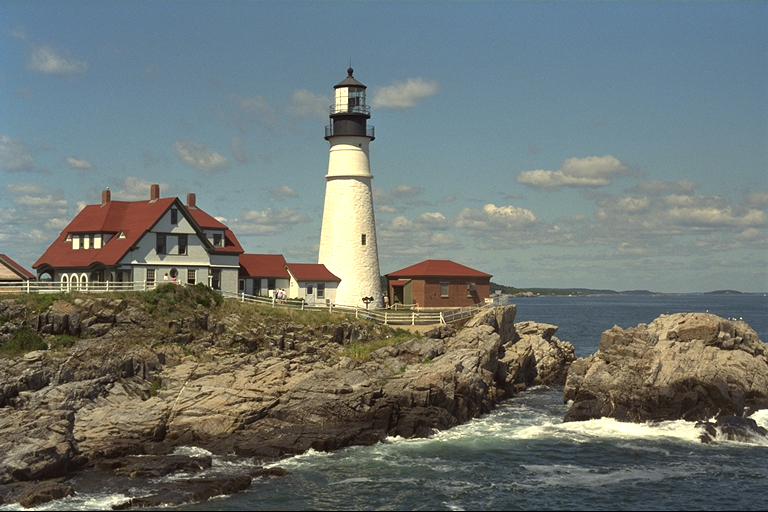}}
\subfigure[Forrest*$\protect\\$   SI =101.405]{\includegraphics[scale=0.076]{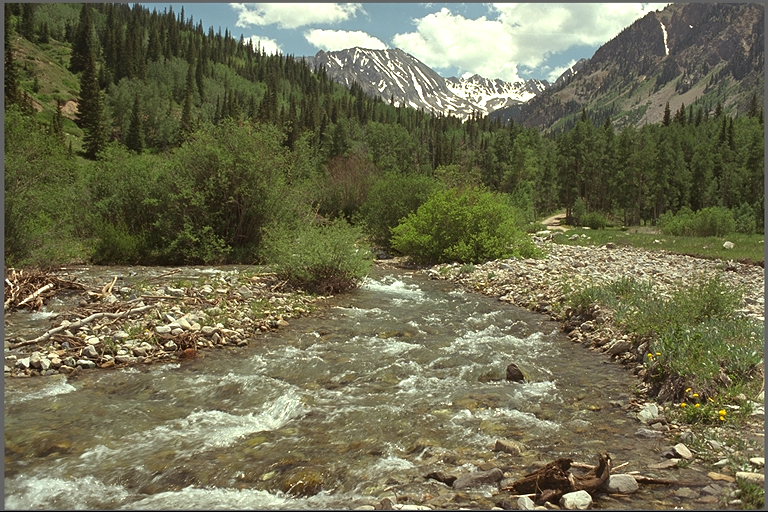}}
\subfigure[Seaspace*$\protect\\$    SI =50.869]{\includegraphics[scale=0.076]{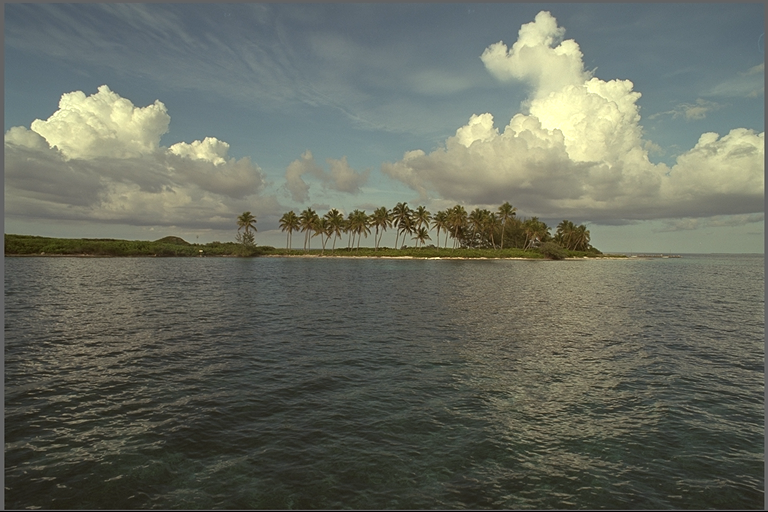}}
\caption{ Images used for subjective assessment from \textbf{Kodak} dataset: 6 images for testing, 2 images for training (star)}
\label{fig:kodak_source_data}
\end{figure*}


\textbf {Additional Studies for Distortion Metric} To explore the impact of various distortion  metrics on the compression efficiency (particularly at low bitrate), we choose MSE, SSIM, MS-SSIM and MSE in feature domain based on VGG19~\cite{simonyan2014very} (MSEf) to conduct extra experiments. For fair comparison, we control the bitrate for each metric as close as possible and optimize the compression framework individually. We find that MSE always shows the over-smoothed results and SSIM provides a strong contrast but a slight brightness shift. For feature domain loss MSEf, it usually causes some noise around texture. Note that we extract the 5th, 10th and 17th layer of VGG19 net as the representations for the shallow-, mid-, high-level features, respectively. The result is shown in Fig.~\ref{fig:kodim19_metric}. 

\section{Subjective Quality Assessment}\label{sec:sqa}
Existing objective quality metrics, \eg, PSNR, SSIM, sometimes are not always consistent with our subjective perception. A straightforward way is to perform the subjective assessment. Towards this goal, we have invited 43 volunteers to participate the assessments. These subjects are from diverse majors including 19 males and 24 females, aging from 18 to 30. All of them have normal vision(or after  correction) and color perception, and almost all of them ($\approx$98\%)  are naive without expertises in video/image compression field.

\begin{figure}[t]
\centering
\subfigure[Boat]{\includegraphics[scale=0.43]{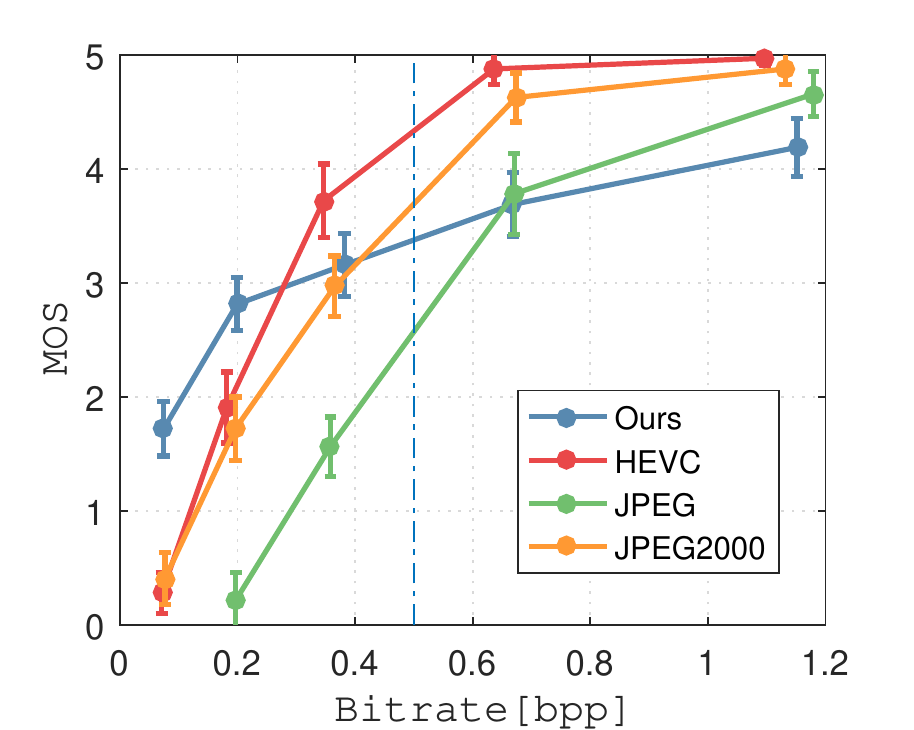}}
\subfigure[Door]{\includegraphics[scale=0.43]{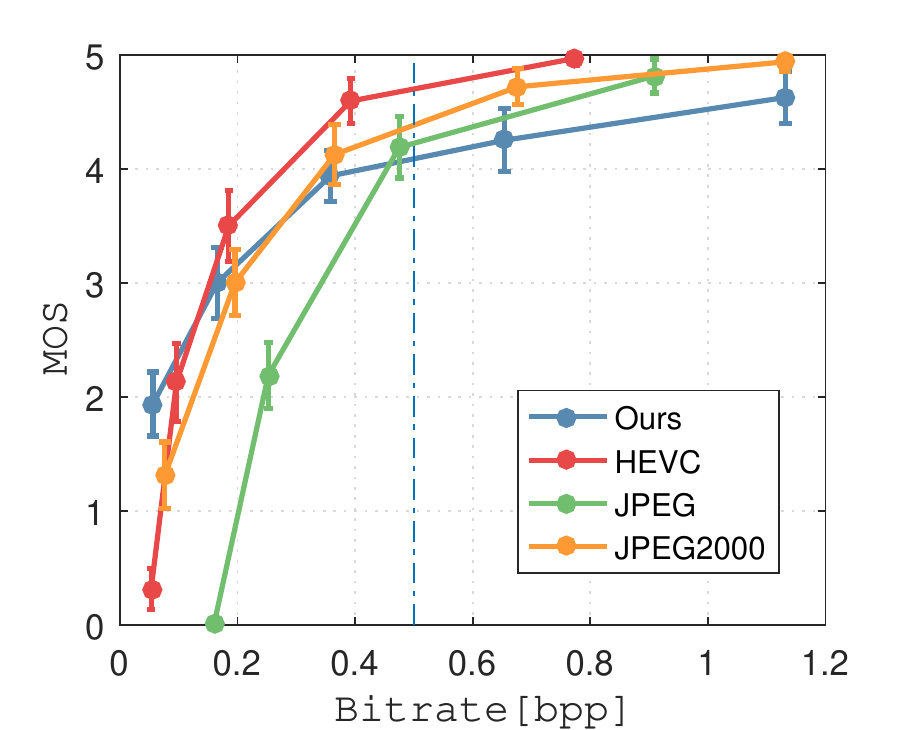}}
\subfigure[Girl]{\includegraphics[scale=0.43]{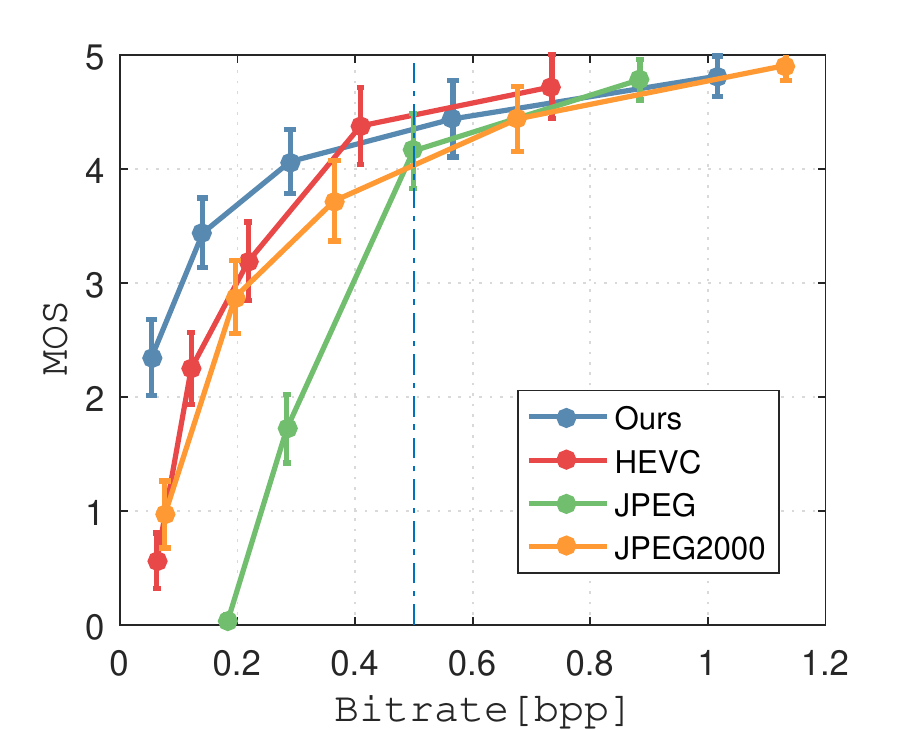}}
\subfigure[Lighthouse]{\includegraphics[scale=0.43]{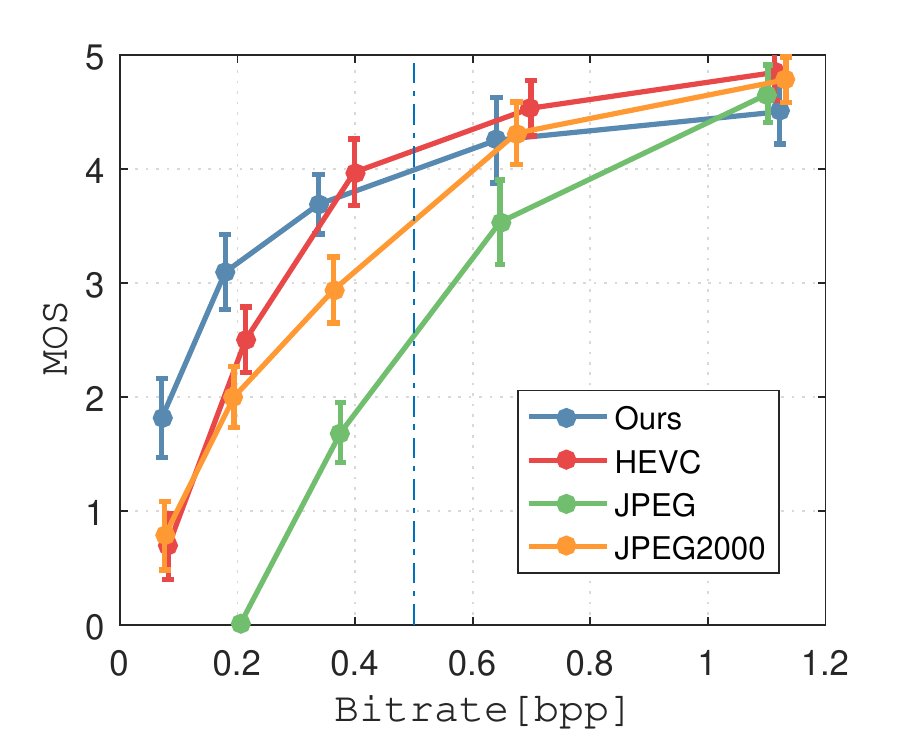}}
\subfigure[Motobike]{\includegraphics[scale=0.43]{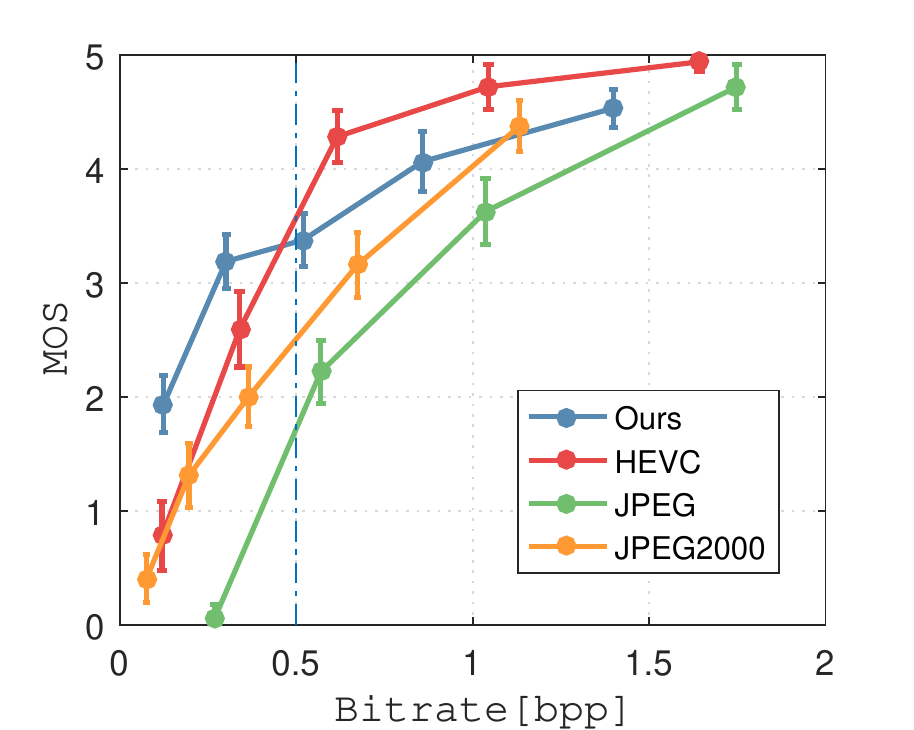}}
\subfigure[Apartment]{\includegraphics[scale=0.43]{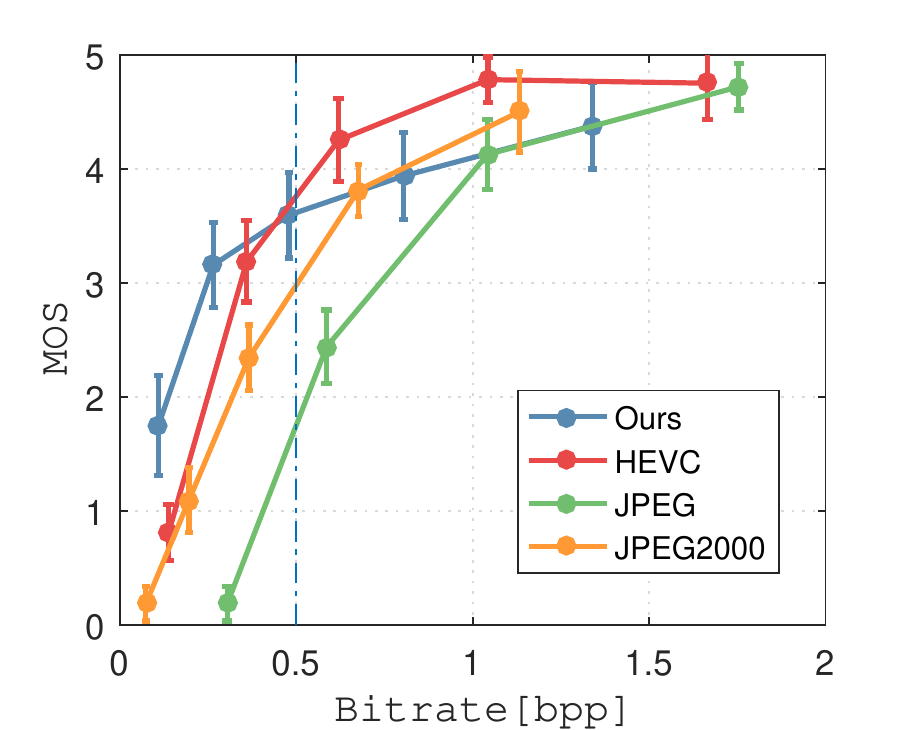}}
\caption{ Illustration of MOS versus bitrate for test contents encoded with all four codecs. The blue vertical line represents the boundary between the low and high bitrates. }
\label{fig:results_sqm}
\end{figure}

\subsection{Test Sequence Pool}
Eight images, all sampled at 768$\times$512 and shown in Fig.~\ref{fig:kodak_source_data}),
are selected from the \textbf{Kodak} dataset for assessments.
These images covers a wide range of content complexity measured by the spatial information index (SI).
Each image was encoded with four codecs, i.e., Ours, BPG(4:4:4), JPEG, JPEG2000, using multiple bitrate settings for sufficient quality scales. Even though we enforce the same bitrates for all contents, it still present the variations for actual encoded bits because of the distinct content characteristics.


\subsection{Test Protocol}
All test sequences are displayed at the center of a 23.8 inch DELL U2414H monitor with a resolution of 1920$\times$1080. Viewers usually rate the image from a 65-cm distance away,  which keeps a $18.2^\circ$ visual angle that ensures the viewer with the most accurate sensation of the image quality. The displayed contents are surrounded with a mid grey level background as recommended by BT.500-13~\cite{BT500-13}.

We adopt the Double Stimulus Impairment Scale (DSIS~\cite{BT500-13}) Variant $\uppercase\expandafter{\romannumeral1}$ to evaluate the impairments across different compression methods.
In this test, each stimulus consists of a pair of images with identical content, while one is the original image as the reference, another is the one with compression noise.
Image pairs are displayed sequentially for about 6 seconds with a 3-seconds pause followed for subjects to give the ratings.

We divide the reference images and their corresponding variants into two sub-groups  to avoid unexpected rating
noise caused by dizziness and tiredness due to a longer duration of assessment. Image pairs with various bitrates but same content are randomly permuted. Different contents occur alternatively in case that the same reference image with its variants are presented for two successive occasions with the same or different levels of impairment. Each image repeats twice in general. We prepare the training sequence to let viewers familiarize with the test protocol before the formal test.

\subsection{Test Analysis}
For each image pair, raw scores from all participants are collected.
 A standard outliers screening is conducted to remove subjects whose ratings deviate from others largely, under the guidelines recommended in~\cite{BT500-13}.
 After data screening, each test sequence has 20 valid ratings. We average all users' ratings for each processed image sample as its mean opinion score (MOS).
 Then we plot the MOS together with confidence interval (CI) versus bitrates in Fig.~\ref{fig:results_sqm}.

 It is observed that the subjective results are reliable as the 95\% confidence interval is relatively small.
Apparently, our method demonstrates better perceptual quality at low bitrate for almost all test images, i.e., $\leq$ 0.5 bpp. For those extremely low bitrates ($\leq$ 0.2 bpp), our method still presents high-quality
and conformable reconstructions, but others often produce very bad image quality with color distortion, blockiness and oversmoothed area, as exemplified in Fig.~\ref{fig:kodim15_comparison}.

\begin{figure}[t]
\centering
\includegraphics[scale=0.53]{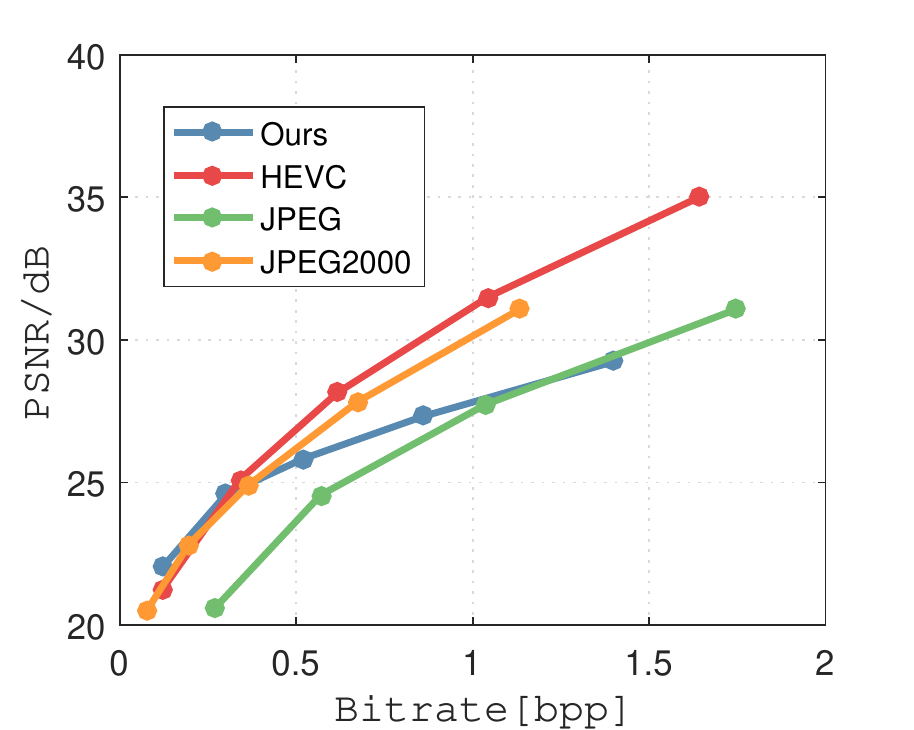}
\caption{Illustration of PSNR versus bitrate for \textit{Motobike}}
\label{fig:psnr_rate}
\end{figure}

For high bitrate scenarios, we have found that BPG and JPEG2000 present better visual quality. One main reason is that MS-SSIM metric used in our method is not favored at high bitrate for end-to-end learning, but rather the PSNR. Figure~\ref{fig:psnr_rate} plots the PSNR versus bitrate for \textit{Motobike} image. We can see that the PSNR of our learning based method is even worse than JPEG at high bitrate case. Hence, it calls for an effective distortion metric that can preserve the fidelity for entire bit rate range when applied in learning based methodology.

\section{Conclusion}
In this paper, we proposed a novel gated context model with embedded priors for end-to-end optimized image compression
and achieve the state-of-the-art performance measured by MS-SSIM and bitrate.
A three separable stacks are used to avoid blind spots in conditional probability prediction for entropy rate modeling. In addition, those embedded priors are exploited and fused with latent features to improve the final image reconstruction, via introduced information compensation network. Residual learning with GDN has proved to maintain both convergence speed and performance.

Additional studies have been extended to investigate various distortion metrics in learning framework, such as MSE, SSIM, MS-SSIM, MSEf, \etc, together with the independent subjective quality assessment, to evaluate the efficiency of our proposed methods and other traditional image compression schemes. MS-SSIM used in current work could preserve high-quality reconstructions at low bitrate (i.e., $\leq$0.5 bpp) scenarios.  A better distortion metric is highly desired for learning based compression, to retain the same performance as MS-SSIM for low bitrate range, and as PSNR or MSE  used in BPG, JPEG2000 for high bitrates.

For future study, we can make our context model deeper to improve image compression performance. Parallelization and acceleration is important to deploy the model for actual usage in practice, particularly for mobile platforms. In addition, it is also meaningful to extend our framework for end-to-end video compression framework with more priors acquired from spatial and temporal information.

\newpage
{\small
\bibliographystyle{ieee}
\bibliography{DIC_for_review_final_v2}
}

\end{document}